\documentstyle[preprint,eqsecnum,aps,tighten,floats,epsfig]{revtex}

\begin{document}
\draft
\title{Out-of-equilibrium thermodynamic relations in systems with\\
aging and slow relaxation
}
\author{Mitsuhiro Kawasaki\cite{byline}\\
}
\address{Department of Physics, Kyushu University, Fukuoka 812-8581, Japan
}
\date{\today}
\maketitle
\begin{abstract}
The experimental time scale dependence of thermodynamic relations in 
out-of-equilibrium systems with aging phenomena is investigated theoretically 
by using only aging properties of the two-time correlation functions and 
the generalized fluctuation-dissipation theorem (FDT). 
We show that there are two experimental time regimes characterized by 
different thermal properties. In the first regime where the waiting time 
is much longer than the measurement time, the principle of minimum work holds 
even though a system is out of equilibrium. In the second regime where 
both the measurement time and the waiting time are long, 
the thermal properties are completely different from properties in 
equilibrium. For the single-correlation-scale systems such as $p$-spin 
spherical spin-glasses, contrary to a fundamental assumption of 
thermodynamics, the work done in an infinitely slow operation 
depends on the path of change of the external field even when the 
waiting time is infinite. On the other hand, for the multi-correlation-scale 
systems such as Sherrington-Kirkpatrick model, the work done in an infinitely 
slow operation is independent of the path. Our results imply that 
in order to describe thermodynamic properties of systems with aging it is 
essential to consider the experimental time scales and history of a system 
as a state variable is necessary.
\end{abstract}

\narrowtext

\section{INTRODUCTION}
\label{sec:introduction}

Glassy systems like spin-glasses and structural glasses
below the glass transition temperatures are out of equilibrium even on the 
macroscopic time scale. Thus, the slow dynamics of glassy systems has been 
a subject of continuous interest in the past years \cite{Vincent96}.
Experimentally, the anomalous dynamical behaviors are characterized by
slow relaxation with long time tail and aging phenomena. 

Among them, aging phenomena are the most striking dynamical behaviors as 
follows: 
Two-time quantities like the correlation functions $C(t,t')$ explicitly 
depend on the time elapsed after the quench $t'$ (the waiting time). 
If the waiting time $t'$ is of microscopic time scale, 
the phenomena are merely transient on the way of relaxation to equilibrium. 
However, dependence on the waiting time continues 
even when $t'$ is so large that 
one-time quantities like the magnetization are 
asymptotically close to time-independent values \cite{Lundgren83}. 
Since these phenomena mean that the dynamics is not stationary, 
aging is a sign showing that these systems are out of equilibrium even in 
macroscopic time scale, i.e. several days or weeks. 

These aging phenomena appear also in mean-field models of spin-glasses
and do not disappear even in the infinite waiting time limit 
\cite{Cugliandolo93,Cugliandolo94}. 
In addition, the phase space of these models decomposes into a
large number of areas separated with infinitely high free energy barriers. 
Thus, glassy systems never reach true 
equilibrium and hence they are beyond the scope of thermodynamics and 
equilibrium statistical mechanics. 
Hence, in order to describe thermodynamic properties of glassy systems, 
out-of-equilibrium thermodynamics based on a dynamical description
without the assumption of ergodicity is necessary. 

In investigation of such anomalous dynamical behavior of glassy systems, 
it was found that aging phenomena have some universal properties. 
Theoretical analyses  
have suggested that there are two time regimes characterized by different
dynamical properties of the two-time correlation function $C(t,t')$ and the
associated linear response function $R(t,t')$ 
\cite{Cugliandolo93,Cugliandolo94}. 
In the first time regime where the time difference $t-t'$ 
is short compared to $t'$, the dynamics looks stationary and the usual
fluctuation-dissipation theorem (FDT) holds. On the other hand, 
in the second time regime where the time difference is comparable to
$t'$, aging phenomena occur, i.e. 
$C(t,t')$ depends on $t'$ apart from dependence on $t-t'$. 
In addition, it is known that the usual FDT 
between the correlation $C(t,t')$ and the response function $R(t,t')$ 
should be modified in a well-defined way which involves the rescaling of 
the temperature \cite{Cugliandolo97}. The modification was found to be 
valid not only for mean-field models but also for other glassy systems: 
spin-glass models with finite-range 
interactions \cite{Franz95}, real spin-glasses \cite{Cugliandolo99}, 
structural glasses \cite{Parisi97,Barrat99,Grigera99} 
and a model of phase separation \cite{Berthier99}. 
In addition, it is known for some glassy systems 
that the correlation function $C(t,t')$ obeys 
the scaling law that it depends on $t$ and $t'$ only through the value of 
$\xi(t)/\xi(t')$, where $\xi(t)$ is a system-dependent increasing function
 of time. 

Aging of the correlation function and the modification of FDT 
imply that properties of the work done by 
modulating an external field in an isothermal process 
are completely different from properties predicted by traditional 
thermodynamics. In addition, the existence of 
the two time scales implies that 
thermodynamic properties must strongly depend on experimental time scales. 
Hence, the experimental time scale dependence of 
the thermodynamic properties of glassy systems should be investigated to
construct out-of-equilibrium thermodynamics for glassy systems.

In order to describe our results precisely, 
we summarize thermodynamics for an isothermal process.
Thermodynamics tells that when one quasistatically changes the 
external field the work needed for the change is independent of a path 
of changing the external field. In addition, the quasistatic work is 
equal to the change of the Helmholtz free energy. 
When the process is not quasistatic, the work is 
larger than the quasistatic work. This fact is called the principle of 
the minimum work and is derived from the second law of thermodynamics. 

We show that the properties described above 
do not hold in systems with aging. 
More precisely, there are two experimental time regimes 
characterized by different thermal properties. 
The first regime is a time-domain where the waiting time is much longer than 
the time lapse of the process. We call the time lapse the measurement time. 
In this regime, the principle of the minimum work 
holds even though a system is out of equilibrium. More precisely, 
when the process is not infinitely slow, the work needed for the process is 
larger than the work for an infinitely slow process. In addition, value of the 
work for the infinitely slow process depends only on the initial 
state and the final state and hence it can play a role of a free energy.

The second regime is the experimental time regime where 
the length of the measurement time are comparable to that of the waiting time. 
In this regime, for the single-correlation-scale
systems such as $p$-spin spherical spin-glasses 
the work done in an infinitely slow operation depends on the 
path of changing the field even when the waiting time is infinite. 
This property form a striking contrast to the consequence of traditional 
thermodynamics described above. On the other hand, 
for the multi-correlation-scale systems such as Sherrington-Kirkpatrick
model, the work done in an infinitely slow operation 
is independent of the path. 

In Sec.\ \ref{section2}, we describe an isothermal process considered in 
this paper and introduce the two time regimes which characterize the 
experimental time scales and play a significant role in this paper. 
In Sec.\ \ref{section3}, we see that in the first time regime, usual 
thermodynamic relations hold even though the system is out of equilibrium. 
The only difference is that the value of the work in an 
infinitely slow operation $W_s$ is different from that of the change of the 
free energy calculated from equilibrium statistical mechanics. 
In Sec.\ \ref{section4}, we present general discussion on 
properties of the work $W_s$ in an infinitely slow operation in the 
second time regime and derive conditions when $W_s$ depends on the path of 
changing the external field. In Sec.\ \ref{section5}, 
by using the results obtained in the previous section, we show that 
for the single-correlation-scale systems the work in an infinitely slow 
operation depends on the path of changing the external field 
as a consequence of aging. Possibility of observation of this path-dependence 
is also discussed. In addition, we show that for the multi-correlation-scale 
systems $W_s$ is not path-dependent. Our results are summarized in 
Sec.\ \ref{section6}, where implications of our results on thermodynamics of 
glassy systems and experimental protocols to observe quasiequilibrium 
properties are discussed.

\section{An isothermal process and two time regimes}
\label{section2}

\begin{figure}
\centerline{\epsfxsize=10cm
\epsffile{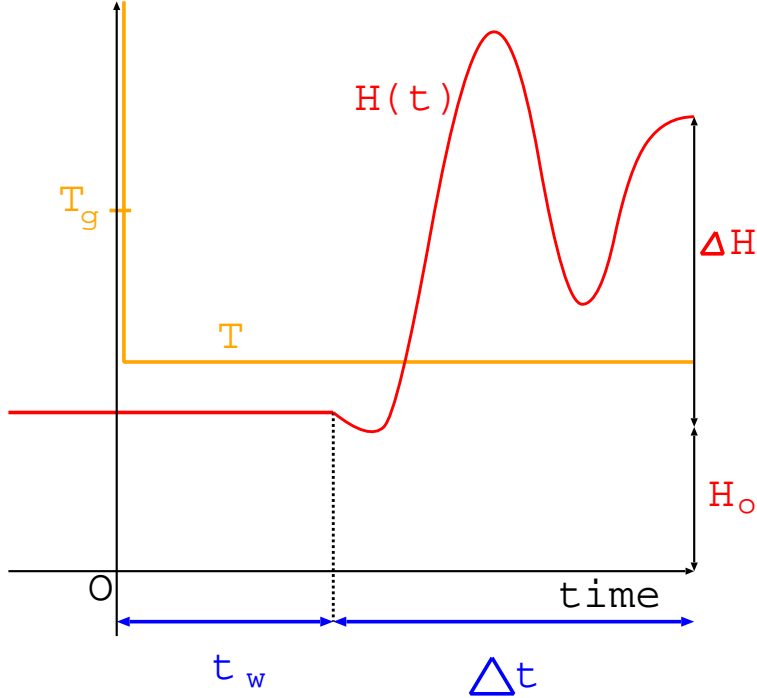}}
\caption{The isothermal process to consider thermodynamic properties 
when aging occurs. The temperature of the heat bath is decreased in a small 
field $H_0$ to a sub-critical temperature at time 0. 
After a waiting time $t_{w}$, the field begins to change according to a 
given time-dependence $H_0+H(t)$. This change of the field continues for 
a period of $\triangle t$. 
The waiting time $t_w$ and the measurement time $\triangle t$ characterize 
the experimental time scales.}
\label{protocol}
\end{figure}

We describe an isothermal process to consider thermodynamic 
properties when aging occurs. 
A simple way to observe aging phenomena is through the following 
field cooling process (Fig.\ \ref{protocol}); 
The temperature of the heat bath is decreased in a small field $H_0$ 
to a sub-critical temperature at time 0. 
After a waiting time $t_{w}$ 
the field begins to change according to a given time-dependence $H_0+H(t)$.
This change of the field continues for a period of $\triangle t$. 
The initial and final values of $H(t)$ are $H(t=t_w) = 0$ and 
$H(t=t_{w}+\triangle t) = \triangle H$. 
We refer to $\triangle t$, the time lapse of the change of the field, 
as the measurement time.
The waiting time $t_{w}$ and the measurement time $\triangle t$ characterize 
the experimental time scales.

When the field is so weak that the response is linear, 
the work $W$ done on the sample during the process 
is given in terms of the response function $R(t,t')$ by 
\begin{equation}
W = - \int_{t_{w}}^{t_{w}+\triangle t} dt \frac{dH(t)}{dt} M(t),
\end{equation}
where the ``magnetization'' $M(t)$ is given by 
\begin{equation}
M(t)=\int_{-\infty}^0 dt' R(t,t') H_0+\int_0^t dt' R(t,t')H_0+
\int_{t_w}^t dt' R(t,t')H(t').
\label{m}
\end{equation}
Since we are interested in long waiting time behavior, the contribution
of the first term of Eq.\ (\ref{m}) is ignored.
For feasibility of showing long time behavior, 
we rewrite Eq.\ (\ref{m}) by integration by part to express $M(t)$ 
in terms of a susceptibility $\chi(t,t')$ 
instead of the response function;
\begin{eqnarray}
W & = & -H_0\int_{t_w}^{t_w+\triangle t} dt \frac{dH(t)}{dt}\chi(t,0)
\nonumber \\
& & - \int_{t_{w}}^{t_{w}+\triangle t} dt \frac{dH(t)}{dt} 
\int_{t_{w}}^{t} dt' \frac{dH(t')}{dt'} \chi(t,t')
\label{w_chi}, 
\end{eqnarray}
where the susceptibility is defined as 
\begin{equation}
\chi(t,t') \equiv \int_{t'}^{t} dt'' R(t,t'') \label{chi}.
\end{equation}

In order to discuss the dependence of the work on the waiting time and 
the measurement time, we rewrite the expression of the work by 
the transformations: $t\rightarrow s \equiv (t-t_{w})/\triangle t$, 
$H(t) \rightarrow h(s) \equiv H(s \triangle t+t_{w})/\triangle H$. 
We assume that $dh(s)/ds$ and $d^2 h(s)/ds^2$ are finite, 
in order to exclude the unrealistic cases where 
the speed and the acceleration of changing the field is infinity. 
For example, the path such as $h(s) = \sqrt{s}$ is excluded, 
since $dh(s)/ds$ and $d^2 h(s)/ds^2$ are infinite at $s=0$.

Thus, the work is reduced to 
\begin{eqnarray}
W/(\triangle H)^{2} = -H_0/\triangle H \int_0^1ds \frac{dh(s)}{ds}
\chi(s \triangle t+t_w,0)
\nonumber \\
-\int_{0}^{1}ds_{1} \frac{dh(s_1)}{ds_1} 
\int_{0}^{s_{1}}ds_{2} \frac{dh(s_2)}{ds_2} \nonumber \\ 
\times 
\chi(s_{1}\triangle t+t_{w}, s_{2}\triangle t+t_{w}) \label{w_s},
\end{eqnarray}
which implies that the dependence of the work on the experimental time 
scales is determined by that of the susceptibility.
Since the first term of Eq.\ (\ref{w_s}) becomes a constant 
$-H_0 \chi(\infty,0)/\triangle H$ in the long waiting time limit 
which we are interested in, 
we will analyze properties of the second term of Eq.\ (\ref{w_s}), i.e. 
the work done in the zero-field cooling process ($H_0=0$), in 
the rest of this chapter for simplicity; 
\begin{equation}
W/(\triangle H)^{2} = -\int_{0}^{1}ds_{1} \frac{dh(s_1)}{ds_1} 
\int_{0}^{s_{1}}ds_{2} \frac{dh(s_2)}{ds_2} 
\chi(s_{1}\triangle t+t_{w}, s_{2}\triangle t+t_{w}) \label{w_s_2}.
\end{equation}

In order to discuss behavior of the susceptibility when aging occurs, 
we recapitulate the long-time behavior of the response function referred to 
in the previous section. It is known that there are two time regimes 
characterized by different behavior of the correlation function $C(t,t')$ and 
the FDT \cite{Cugliandolo97};
\begin{itemize}
\item At long times $t$ and $t'$ such that $t-t' \ll t'$, the correlation 
function is the function of only the time difference $t-t'$, i.e. the 
time-translational invariance (TTI) holds. In addition, although the 
sample is out of equilibrium, the usual FDT holds \cite{FDT} as 
\begin{equation}
R(t-t') = \frac{1}{k_B T}\frac{\partial C(t-t')}{\partial t'},
\end{equation}
where $C(\tau)$ is defined as $\lim_{t \rightarrow \infty} C(\tau+t,t)$. 
Since the properties of two-time quantities in this time regime are the 
same as that in equilibrium, this time regime is called the quasiequilibrium 
regime.
\item Whereas, at long and well-separated times such that $t-t' \sim t'$, 
aging occurs, i.e. the two-time correlation function $C(t,t')$ depends on $t'$ 
even in the long time limit $t' \rightarrow \infty$. This implies no TTI and 
this time regime is called the aging regime. In addition, it is known that 
in the aging regime the FDT is modified as 
\begin{equation}
R(t,t') = \frac{X[C(t,t')]}{k_B T}\frac{\partial C(t,t')}{\partial t'},
\label{generalized_FDT}
\end{equation}
where the FDT violation factor $X$ is a function which depends on $t,t'$ only 
through the dependence of the correlation function $C(t,t')$ 
\cite{Cugliandolo94}. Thus, a system cannot be considered to be in a 
quasiequilibrium state, since the usual FDT is strongly violated. 
When only one correlation scale exists apart from the quasiequilibrium regime, 
it is known that the FDT violation factor is a constant.
\end{itemize}

In order to clarify the meaning of the time region of the aging regime
($t-t' \sim t'$), we give an explicit expression of aging of the
correlation function as 
\begin{equation}
\lim_{{\tau\rightarrow\infty}\atop{\; \mu=t'/g(\tau)}} C(\tau+t',t')=
\lim_{\tau\rightarrow\infty}C[\tau+\mu g(\tau),\mu g(\tau)] \equiv 
\hat{C}(\mu),
\label{explicit_expression_aging}
\end{equation}
where $g(t)$ is a system-dependent function which characterizes the aging 
regime and $\tau$ is the time difference $t-t'$. 
The correlation function $\hat{C}(\mu)$ depends on the waiting time $t'$
through the value of $\mu$. 
Occurrence of aging means that the
limiting function $\hat{C}(\mu)$ takes a non-trivial value such that
$0<\hat{C}(\mu)<q$. Here, $q$ is the dynamical E-A order parameter. Thus, the
time region of the aging regime is the time region 
where $t'/g(t-t')$ is finite. 
Here, we assume that $\mu$ and $g(\tau)$ are positive. 

\begin{figure}
\epsffile{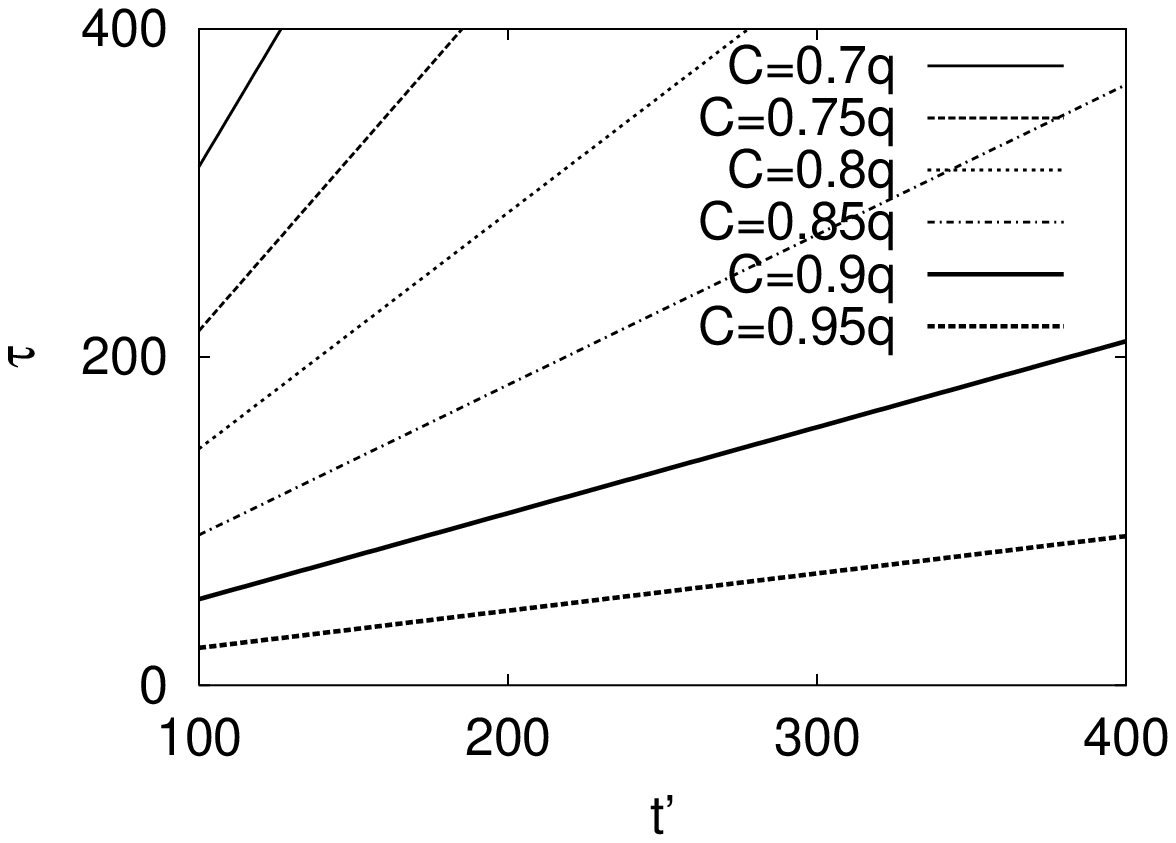} \epsffile{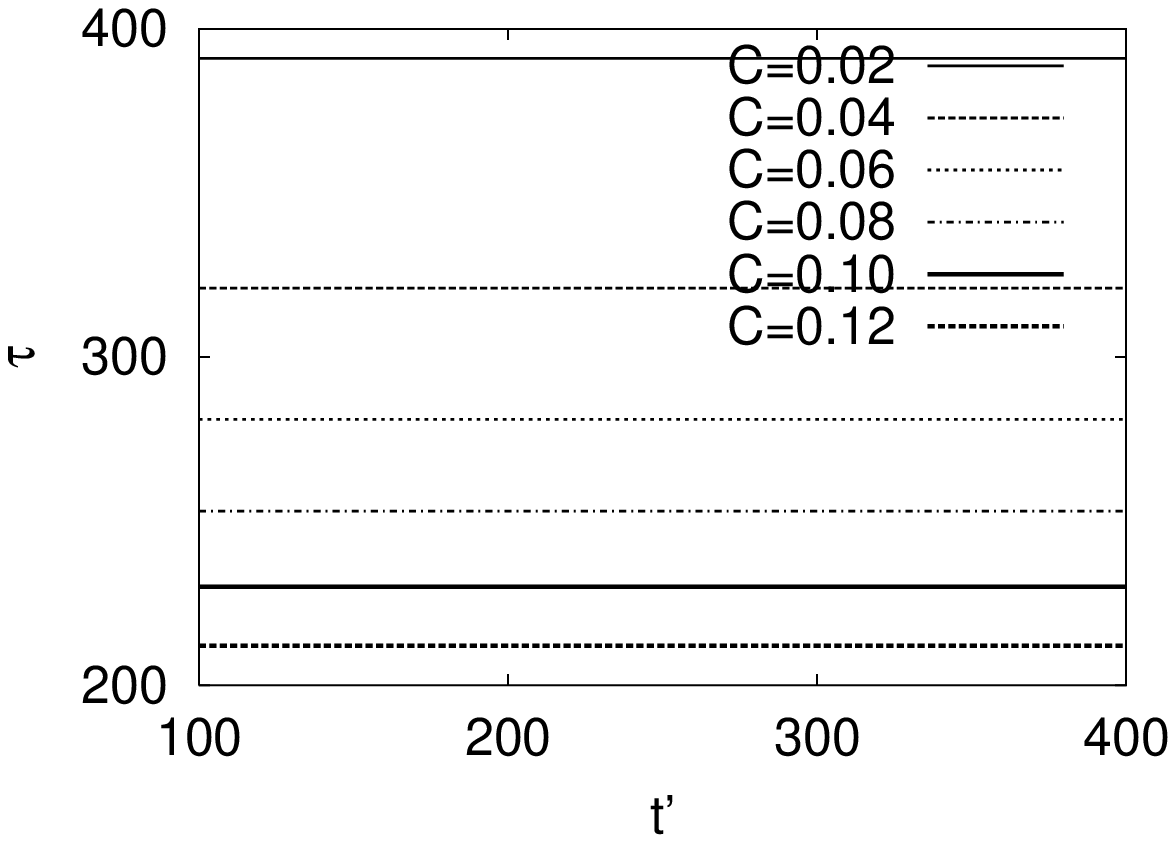}
\caption{
 (a) The contour plot on $\tau-t'$ plain of the correlation function 
 when aging occurs; 
 $C(t,t')=q(t'/t)^{\gamma} \ (\gamma=0.25)$, which is the correlation
 function in the aging regime for the 5-spin spherical spin-glass model.  
 The plot when aging occurs is completely different from that when aging
 does not occur in that the contour lines are not parallel to the
 horizontal axis. This contour plot gives the system-dependent function
 $g(t)$ which characterizes the aging regime as a contour line of the
 correlation function, i.e. $\mu=t'/g(\tau)$. The straight contour lines
 in this figure imply that $g(\tau)$ is a linear function
 of $\tau$.
 (b) The contour plot when aging does not occur; 
 $C(\tau+t',t')=\exp[-\tau/\tau_0]$. Since the
 value of the correlation function does not depend on $t'$, the contour
 lines are parallel to the horizontal axis. }
\label{contour}
\end{figure}

This definition of aging is illustrated in terms of contour plot of the
correlation function on $\tau-t'$ plain (Fig.\ \ref{contour} (a)).
The plot when aging occurs is completely different from that when aging
does not occur (Fig.\ \ref{contour} (b)). 
The contour lines give the system dependent function
$g(t)$ which characterizes the aging regime, since the contour lines are
given by $\mu=t'/g(\tau)$ in the contour plot.

We show two examples of the function $g(t)$ and the aging regime for
systems like spherical spin-glasses and real spin-glasses. In these
systems, it is known that the correlation function behaves as a function 
of $\xi(t)/\xi(t')$, where the scaling function $\xi(t)$ is a
system-dependent function \cite{Vincent96}. The function $g(t)$ is given
in terms of $\xi(t)$ as $g^{-1}(t)=\xi^{-1}[r\xi(t)]-t$, where $r$ is a
constant larger than unity. The inverse function of $\xi(t)$ exists since
$\xi(t)$ is a monotonically increasing function. For example, 
\begin{enumerate}
\item When $\xi(t)=t$,
\begin{equation}
\lim_{{\tau \rightarrow \infty} \atop{\; \mu = t'/\tau}}
      \frac{\xi(\tau+t')}{\xi(t')}=
\lim_{{\tau \rightarrow \infty} \atop{\; \mu = t'/\tau}}
      \frac{\tau+t'}{t'} = \lim_{\tau \rightarrow \infty} 
\frac{\tau+\mu \tau}{\mu \tau}=\frac{1+\mu}{\mu}.
\end{equation}
Thus, 
$\lim_{{\tau \rightarrow \infty} \atop{\; \mu = t'/\tau}} C(\tau+t',t')$ 
is a function of $\mu$ and hence the aging regime is the time regime 
where $t'/\tau$ is finite. 
\item When $\xi(t)=\exp[(t/\tau_0)^{1-\alpha}/(1-\alpha)] (0<\alpha <1)$,
\begin{equation}
\lim_{{\tau \rightarrow \infty} \atop{\; \mu = t'/\tau^{1/\alpha}}} 
\frac{\xi(\tau+t')}{\xi(t')}=\exp(\tau_0^{\alpha-1} \mu^{-\alpha}).
\end{equation}
Thus, 
$\lim_{{\tau \rightarrow \infty} \atop{\; \mu = t'/\tau^{1/\alpha}}} 
C(\tau+t',t')$ is a function of $\mu$ and hence the aging regime is the 
time regime when $t'/\tau^{1/\alpha}$ is finite.
\end{enumerate}

From these results, it is shown by the definition of the susceptibility 
Eq.\ (\ref{chi}) that the behavior of the susceptibility depends on the time 
regimes according to the dependence of the behavior of the response function.
In the quasiequilibrium regime, the susceptibility depends only 
on the time difference 
$\tau \equiv t-t'$ and is given by the correlation function as 
\begin{equation}
\chi(\tau) = -\frac{1}{k_{B}T}[C(\tau)-C(0)] \label{chi_f}
\end{equation}
where $C(\tau)\equiv \lim_{t'\rightarrow\infty}C(\tau+t',t')$. 
Whereas, in the aging regime, time translational invariance (TTI) 
does not hold and the susceptibility is 
given by the correlation function as 
\begin{equation}
\chi(t,t') = \frac{1}{k_{B}T} \int_{C(t,t')}^{C(t,t)} dC X(C) \label{chi_a}
\end{equation}
where $X$ is the FDT violation factor defined by Eq.\ (\ref{generalized_FDT}).
It implies that the susceptibility is a function of the correlation function. 

Since the difference of the two arguments of the susceptibility in 
Eq. (\ref{w_s_2}) is $(s_{1}-s_{2})\triangle t$, 
there are also two time regimes characterized by different dependence of the 
work on the experimental time scales $\triangle t$ and $t_w$.
\begin{itemize}
\item When the waiting time is much longer than the measurement time 
($\triangle t \ll t_{w}$), 
the susceptibility which appears in Eq.\ (\ref{w_s_2}) 
obeys Eq.\ (\ref{chi_f}) 
since $(s_{1}-s_{2})\triangle t \ll s_{2} \triangle t+t_{w}$ holds. 
Hence, from Eqs.\ (\ref{w_s_2}) and (\ref{chi_f}) the work in this regime is 
written as 
\begin{eqnarray}
W/(\triangle H)^2 
& = & -\frac{C(0)}{2k_{B}T} +\frac{1}{k_{B}T} \int_{0}^{1}ds_{1} 
\frac{dh(s_{1})}{ds_{1}}
\nonumber \\
& &  \times \int_{0}^{s_{1}}ds_{2} 
\frac{dh(s_{2})}{ds_{2}} C[(s_{1}-s_{2})\triangle t] \label{w_f}.
\end{eqnarray}
In the derivation, 
\begin{equation}
\int_{0}^{1}ds_{1} \frac{dh(s_{1})}{ds_{1}} 
\int_{0}^{s_{1}}ds_{2} \frac{dh(s_{2})}{ds_{2}}=1/2 \label{integration}
\end{equation}
for any path $h(s)$ is used.

\item On the other hand, when both the waiting time and 
the measurement time are long ($\triangle t \sim t_w$), 
the susceptibility obeys Eq. (\ref{chi_a}) since 
$(s_{1}-s_{2})\triangle t \sim s_{2} \triangle t+t_{w}$ holds 
\cite{time-difference}.
In this regime, it is seen from Eq.\ (\ref{chi_a}) that 
the work is a functional of the correlation function which shows aging. 
The ambiguous relation $t_w \sim \triangle t$ is described explicitly 
 in Sec.\ \ref{section4} with the function $g(t)$ defined in 
Eq.\ (\ref{explicit_expression_aging}).
\end{itemize}

We refer to the former case as the quasiequilibrium regime and 
the latter case as the aging regime without any confusion with the 
time regimes characterized by the behavior of the correlation and the 
response function.

\section{the work in the quasiequilibrium regime}
\label{section3}

In this section, we investigate the properties of the work done on the sample 
when the measurement time $\triangle t$ 
is short compared to the long waiting time $t_{w}$ 
(the quasiequilibrium regime). 

\subsection{The work done in an slow process}

First, we derive the work done in an slow operation such that 
the measurement time $\triangle t$ is so long that the 
correlation function relax to a time-independent value. 
This process is formulated by taking the infinite measurement time 
limit $\triangle t \rightarrow \infty$.
Glassy systems are out of equilibrium even in such an infinitely slow
process. Hence, 
we call a process where the external field is changed infinitely slowly
a ``slow process'' instead of a ``quasistatic process''. 

Since for the slow process the time lapse of the process is 
infinity, the work done in the slow process 
in the quasiequilibrium regime is given by 
taking the limit $\triangle t \rightarrow \infty$ 
after taking the infinite waiting time limit of Eq. (\ref{w_f}). 
Since the integrand, the correlation function, is finite, 
the order of limit and integration can be changed. 
Thus, we can write the work $W_s$ for the slow process as 
\begin{eqnarray}
W_s/(\triangle H)^2 
& = & \lim_{\triangle t \rightarrow \infty} \lim_{t_w \rightarrow \infty} 
W/(\triangle H)^{2} \nonumber \\
& = & -\frac{C(0)}{2k_{B}T}+ \frac{1}{k_{B}T}\
\int_{0}^{1}ds_{1} \frac{dh(s_{1})}{ds_{1}} \nonumber \\ 
& & \times \int_{0}^{s_{1}}ds_{2} \frac{dh(s_{2})}{ds_{2}} 
\lim_{\triangle t \rightarrow \infty} C[(s_{1}-s_{2})\triangle t].
\end{eqnarray}

Here, in order to give the expression of the work $W_s$, 
we introduce the dynamical Edwards-Anderson (E-A) order parameter defined as 
\begin{equation}
q \equiv \lim_{\tau \rightarrow \infty}\lim_{t_{w}\rightarrow \infty}
C(\tau+t_{w},t_{w}).
\end{equation}
Using Eq. (\ref{integration}) and the above definition,  
the work $W_{s}$ for the slow process is given by 
\begin{equation}
W_{s} = -\frac{(\triangle H)^{2}}{2k_{B}T} 
[C(0)-q] \label{w_qs_f} \label{quasistatic_work_f}.
\end{equation}
Therefore, the work for the slow process in the quasiequilibrium regime 
is the difference of a state function, 
since the right hand side of Eq. (\ref{w_qs_f}) 
is independent of the path of changing the field $h(s)$ and dependent only on 
the thermodynamic variables ($\triangle H$ and $T$) and constants intrinsic 
to the system ($C(0)$ and $q$). 
It is important to note that this property is derived by 
using only the FDT.

On the other hand, for usual systems apart from 
glassy systems, one may expect that equilibrium 
is usually achieved when $t_w$ is long ($t_w \rightarrow \infty$). 
So, thermodynamics can be applied and the work for the slow process is equal 
to the change of the Helmholtz free energy, $\triangle F_{eq}$, 
calculated by equilibrium statistical mechanics. 
However, we show below that this naive expectation fails for glassy systems. 
More precisely, the work for the slow process, 
Eq.\ (\ref{quasistatic_work_f}), is different from the change of the Helmholtz 
free energy calculated by equilibrium statistical mechanics. 
It is because the glassy systems are out of equilibrium even when 
the waiting time is infinite.

Assuming that $A(t)$ is a physical quantity coupled to the external field 
$H(t)$ linearly, 
we see from statistical mechanics with the 
assumption of ergodicity that the isothermal susceptibility $\chi_T$ is given
by 
\begin{equation}
\chi_T=\frac{1}{k_B T}
\langle \langle A^2 \rangle_{eq}-\langle A \rangle^2_{eq} \rangle_J
\label{susceptibility_eq}
\end{equation}
where $\langle \cdots \rangle_{eq}$ denotes average over the 
Gibbs-Boltzmann distribution and $\langle \cdots \rangle_J$ denotes 
disorder-average. 
Thus, the free energy difference $\triangle F_{eq}$ due to change $\triangle
H$ of the external field is 
\begin{equation}
\triangle F_{eq} = -\frac{(\triangle H)^2}{2 k_B T}
\langle\langle A^2 \rangle_{eq}-\langle A \rangle^2_{eq}\rangle_J
\label{free_energy_difference}
\end{equation}
where $\langle\langle A \rangle^2_{eq}\rangle_J$ 
is the usual E-A order parameter. 
Since ergodicity is broken for glassy systems, the phase space
decomposes into many pure states. 
Assuming that $\langle \cdots \rangle_a$ denotes the thermal average in local
equilibrium in the pure state $a$ and $F_a$ denotes the free energy at
the pure state $a$, we see that 
\begin{eqnarray}
& \langle A^2 \rangle_{eq} = \sum_a P^{eq}_a \langle A^2 \rangle_a, \\
& \langle A \rangle_{eq}^2 = (\sum_a P^{eq}_a \langle A \rangle_a)^2.
\label{EA_parameter_pure_state}
\end{eqnarray}
where $P_a^{eq}$ is the probability that the system is found at the pure
state $a$ in true equilibrium and is given by 
$P_a^{eq}\equiv \exp(-\beta F_a)/Z$. 

In order to compare $W_s$, Eq.\ (\ref{w_qs_f}), with $\triangle F_{eq}$, 
Eq.\ (\ref{free_energy_difference}), we rewrite Eq.\ (\ref{w_qs_f}) 
in terms of pure states. 
Since 
the correlation function is given by 
$C(t,t')=\langle\langle A(t)A(t') \rangle\rangle_J$ and 
the local equilibrium in 
pure states is achieved in the long time limit, 
\begin{eqnarray}
&& C(0) \equiv \lim_{t'\rightarrow\infty}\langle\langle {A(t')}^2 \rangle
\rangle_J=\sum_a P^{ini}_a \langle\langle A^2 \rangle_a\rangle_J, \\
&& q \equiv \lim_{\tau\rightarrow\infty}\lim_{t'\rightarrow\infty}
\langle\langle A(\tau+t')A(t')\rangle\rangle_J = \sum_a P^{ini}_a 
\langle\langle A \rangle_a^2\rangle_J
\label{q_pure_state}
\end{eqnarray}
where $P^{ini}_a$ is the probability that the system is found in pure
state $a$ at time $0$. 
Since the system is out of equilibrium at time $0$, 
$P_a^{eq} \neq P_a^{ini}$. Thus, $\langle\langle A^2
\rangle_{eq}\rangle_J$ is not equal
to $C(0)$. In addition, from Eqs.\ (\ref{EA_parameter_pure_state}) and 
(\ref{q_pure_state}), $\langle\langle A \rangle_{eq}^2\rangle_J$ 
is not equal to $q$. 
Consequently, we conclude that the work $W_s$ for the slow process does
not coincide with the free energy difference calculated by statistical
mechanics with assumption of ergodicity. 

\subsection{The work when the measurement time is finite}

We discuss the properties of the work 
when the measurement time is finite. 
From Eqs. (\ref{w_f}) and (\ref{quasistatic_work_f}), 
the difference between the work when the process is not slow and the work for
the slow process is given by 
\begin{equation}
W-W_{s} = 
\frac{(\triangle H)^2}{k_{B}T}\int_{0}^{1}ds_{1} \frac{dh(s_{1})}{ds_{1}} 
\int_{0}^{s_{1}}ds_{2} \frac{dh(s_{2})}{ds_{2}} 
\{ C[(s_{1}-s_{2})\triangle t]-q \} \label{devi_work}. 
\end{equation}
$C[(s_{1}-s_{2})\triangle t]>q$ when $\triangle t$ is finite, 
since the quasiequilibrium regime is considered.
Hence, the difference from the work for the slow process 
is positive for any path 
of changing the field, when the measurement time is finite. 
This implies the principle of minimum work;  
\begin{equation}
W \geq W_{s},
\end{equation}
where the equality holds only when the measurement time is infinite,
i.e. the slow-process limit.

Our result is very similar to the consequence of thermodynamics which 
tells the work for 
non-quasistatic process is larger than the change of the Helmholtz 
free energy. However, our result is different from that and 
beyond the scope of thermodynamics 
since in our discussion the initial state and the final state of 
the process are out of equilibrium and the value of the work for the slow
process is different from the value of change in the Helmholtz free energy 
derived by equilibrium statistical mechanics.

\subsection{Long measurement time behavior of $W-W_s$}
\label{difference}

The long measurement time behavior
of $W-W_s$ is analyzed in this subsection. 
We show that the behavior of $W-W_s$ when the measurement time 
$\triangle t$ is long but finite is determined by the long time behavior of 
the correlation function.
We describe below the results for 
four types of behavior of the correlation function which 
include almost all types of relaxation of the correlation, 
e.g., the exponential, the power law \cite{Ogielski85}, 
the logarithmic \cite{Ocio86} and the stretched exponential relaxation 
\cite{Ngai79}. 
The derivations are given in Appendix\ \ref{appendix}.

\begin{enumerate}

\item When the correlation function behaves as 
$\lim_{\tau \rightarrow \infty}\tau[C(\tau)-q]=0$, 
the difference is given by 
\begin{equation}
W-W_{s}\simeq \frac{(\triangle H)^{2}}{k_{B}T}K_{\infty}\int_{0}^{1}ds 
\left| \frac{dh(s)}{ds} \right|^{2} \frac{1}{\triangle t}, 
\end{equation}
where 
\begin{equation}
K_{\infty}\equiv \int_{0}^{\infty} d\tau [C(\tau)-q] \label{k}
\end{equation}
and $W_{s}$ denotes the work for the slow process. 
This case includes the power law relaxation such that 
$C(\tau)\simeq q+c \tau^{-\alpha}$ when $\alpha > 1$ and 
the stretched exponential relaxation ($\exp(-a \tau^{n}), 0<n<1$) 
as well as the exponential relaxation \cite{complementality}.

\item When the correlation function behaves as 
$\lim_{\tau \rightarrow \infty}\tau[C(\tau)-q]=c$, 
the difference is given by 
\begin{equation}
W-W_{s}\simeq \frac{(\triangle H)^{2}}{k_{B}T} c \int_{0}^{1}ds 
\left| \frac{dh(s)}{ds} \right|^{2} \frac{\ln(\triangle t)}{\triangle t}. 
\end{equation}
These two results show that $W-W_{s}$ is proportional to the inverse of the 
measurement time only when $\lim_{\tau \rightarrow \infty}\tau[C(\tau)-q]=0$. 
In these two cases, the difference from the work for the slow process 
takes the minimum value when the field increases linearly as $h(s)=s$, 
since we assume that $dh(s)/ds$ and $d^2 h(s)/ds^2$ is finite.

\item When the correlation function obeys the power law relaxation, such that 
$C(\tau)\simeq q+c \tau^{-\alpha}$ ($0 < \alpha < 1$), 
the difference is given by 
\begin{eqnarray}
W-W_{s} & \simeq & \frac{(\triangle H)^{2}}{k_{B}T} c 
\int_{0}^{1} ds_{1} \frac{dh(s_{1})}{ds_{1}} 
\nonumber \\
& & \times \int_{0}^{s_{1}} ds_{2} \frac{dh(s_{2})}{ds_{2}} 
\frac{1}{(s_{1}-s_{2})^{\alpha}} \frac{1}{(\triangle t)^{\alpha}} 
\label{power}.
\end{eqnarray}
In this case, the difference from $W_s$ obeys the power 
law whose exponent is equal to the exponent of the correlation function.

\item When the correlation function obeys the 
logarithmic relaxation $C(\tau) \simeq q+c/\ln(\tau)$, 
the difference also obeys the logarithmic relaxation as 
\begin{equation}
W-W_{qs} = \frac{(\triangle H)^{2}}{2 k_{B}T} \frac{c}{\ln(\triangle t)} 
+ O[\frac{\ln(\ln\triangle t)}{(\ln\triangle t)^{2}}]
\label{log}.
\end{equation}
In this case, the difference does not depend on the path $h(s)$.
\end{enumerate}

Experimentally, these results tell 
how long the measurement should take 
and the suitable path of changing the external field 
in order to determine the value of the work for the slow process, 
i.e. the difference of a state function in quasiequilibrium regime.

\section{The work in the aging regime; general results}
\label{section4}

In this section, we discuss the properties of the work in
the isothermal slow process when both the waiting time and the measurement
time are long, i.e. in the aging regime, by using the modified FDT and
aging of the correlation function introduced in Eqs.\ 
(\ref{generalized_FDT}) and (\ref{explicit_expression_aging}).
Condition when the work for the slow process depends on the 
path changing the external field is obtained. 
By using the results obtained in this section, 
the path-dependence of the work for particular systems is 
discussed in the next section, Sec.\ \ref{section5}.

\subsection{Path-dependence of the work for the slow process}
\label{subsection1}

As shown in Sec.\ \ref{section2}, in the aging regime the work is a
functional of the correlation function which shows aging. 
The slow process of the aging regime is given by taking the long 
measurement time limit $\triangle t \rightarrow \infty$ with 
holding the relation $t_w \sim \triangle t$, which guarantees that 
the slow-process limit is taken within the aging regime. 
The slow-process limit has always this meaning in this section.

The order of the slow-process limit and the integration can 
be changed since the susceptibility $\chi$ is finite. 
From Eq.\ (\ref{w_s_2}) the work $W_{s}$ for the slow process 
in the aging regime is given by 
\begin{eqnarray}
W_{s}/(\triangle H)^2 & = & -\int_{0}^{1}ds_{1} \frac{dh(s_{1})}{ds_{1}} 
\int_{0}^{s_{1}}ds_{2} \frac{dh(s_{2})}{ds_{2}} 
\nonumber \\
& & \times
\chi [ \lim_{{\triangle t \rightarrow \infty} \atop{\; t_w \sim \triangle t}} 
C(s_1 \triangle t+t_w,s_2 \triangle t+t_w)]
\label{qs_w_a}. 
\end{eqnarray}
Thus, the dependence of $\lim_{{\triangle t \rightarrow \infty} \atop{\; t_w \sim \triangle t}} C(s_1 \triangle t+t_w,s_2
\triangle t+t_w)$ on $s_1$ and $s_2$ determines the dependence of the 
work for the slow process on the path of changing the field $h(s)$; 
$W_s$ is independent of $h(s)$ if 
$\lim_{{\triangle t \rightarrow \infty} \atop{\; t_w \sim \triangle t}} C(s_1 \triangle t+t_w,s_2 \triangle t+t_w)$ does not depend on
$s_1$ and $s_2$. On the other hand, $W_s$ is a functional
of $h(s)$ if $\lim_{{\triangle t \rightarrow \infty} \atop{\; t_w \sim \triangle t}} C(s_1 \triangle t+t_w,s_2 \triangle t+t_w)$ depends on 
$s_1$ and $s_2$. We derive conditions when $W_s$ depends on $h(s)$ 
in the rest of this section.

In order to analyze $\lim_{{\triangle t \rightarrow \infty} \atop{\; t_w \sim \triangle t}} C(s_1 \triangle t+t_w,s_2
\triangle t+t_w)$, we clarify the meaning of $t_w \sim \triangle t$ or 
the aging regime by using $g(\tau)$ in Eq.\ (\ref{explicit_expression_aging}). 
By substitution $t' \rightarrow s_2 \triangle t+t_w, 
\tau \rightarrow (s_1-s_2) \triangle t$ in Eq.\ 
(\ref{explicit_expression_aging}), 
we see that the following equation holds in the region except for 
the point $s_1=s_2$ which does not contribute to the value 
of the work by itself;
\begin{equation}
\lim_{{\triangle t \rightarrow \infty} \atop{\; 
\mu = \frac{s_2 \triangle t+t_w}{g[(s_1-s_2)\triangle t]}}} 
C(s_1 \triangle t+t_w,s_2\triangle t+t_w) = \hat{C}(\mu).
\label{2}
\end{equation}
This equation implies that the aging regime is the experimental time
regime where $(s_2 \triangle t+t_w)/g[(s_1-s_2)\triangle t]$ is finite
and the slow-process limit of the correlation function in 
the integrand in Eq. (\ref{qs_w_a}) is a function of 
$\lim_{{\triangle t \rightarrow \infty} \atop{\; t_w \sim \triangle t}} 
(s_2 \triangle t+t_w)/g[(s_1-s_2)\triangle t]$.

\begin{figure}
\centerline{\epsfxsize=10cm \epsffile{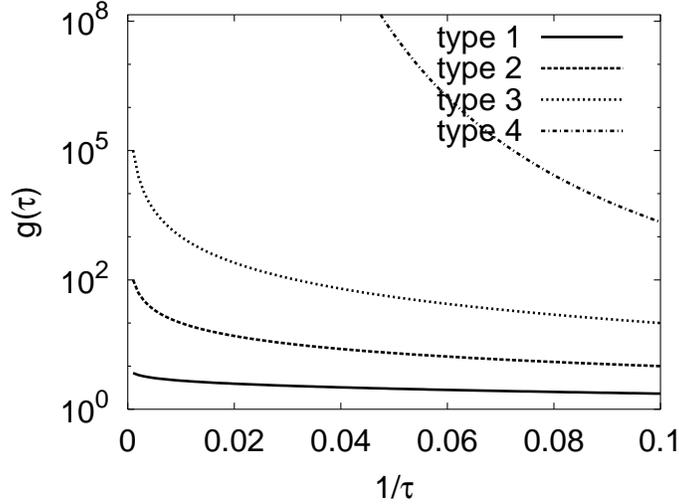}}
\caption{The schematic illustration of the four types of the 
long-time behavior of the function $g(\tau)$. For example, 
$g(\tau)=\log \tau$ (type 1.), $g(\tau)=\tau$ (type 2.), 
$g(\tau)={\tau}^2$ (type 3.), 
$g(\tau)=\exp{\tau}$ (type 4.) are plotted against $1/\tau$. 
It is important to note that these four types exhaust all possibilities of 
long-time behavior.}
\label{g}
\end{figure}

We discuss below the aging regimes and dependence of 
$\lim_{{\triangle t \rightarrow \infty} \atop{\; t_w \sim \triangle t}} C(s_1 \triangle t+t_w,s_2 \triangle t+t_w)$ on $s_1$ and $s_2$ 
in four cases of the different long-time behaviors of $g(t)$ 
which exhaust all possibilities (see Fig.\ \ref{g}).

\begin{enumerate}
\item When $\lim_{\tau\rightarrow\infty}\tau/g(\tau)=\infty$, 
$t'/g(t-t')$ is finite 
only in the region $t' \ll t-t'$. 
In the region of integration of Eq.\ (\ref{qs_w_a}) except for the point 
$s_2=0$, 
\begin{equation}
(s_2\triangle t+t_w)/g[(s_1-s_2)\triangle t]\rightarrow\infty
\label{aging_infty}
\end{equation}
as $\triangle t\rightarrow\infty$ since $t_w > 0$.
An example of such $g(\tau)$ is $g(\tau)=\log \tau$. 
When $g(\tau)=\log \tau$ and $s_2 \neq 0$, 
$\lim_{\triangle t\rightarrow\infty}(s_2\triangle t+t_w)/g[(s_1-s_2)
\triangle t]=\infty$ as shown in Eq.\ (\ref{aging_infty}). 
Since the point $s_2=0$ does not contribute to the value of the work by
      itself, the experimental time regime where $(s_2\triangle t+t_w)/
g[(s_1-s_2)\triangle t]$ is finite does not exist in this case. 
If there is no other aging regime except for that 
characterized by such $g(\tau)$ exists, only the
      quasiequilibrium regime contributes to the value of the work and
      hence the work for the slow process is independent of $h(s)$ and 
is given by Eq.\ (\ref{quasistatic_work_f}).

\item When $g(\tau)=\tau$, 
\begin{equation}
\frac{s_2 \triangle t+t_w}{g[(s_1-s_2)\triangle t]} = 
\frac{s_2 \triangle t+t_w}{(s_1-s_2)\triangle t} = 
\frac{s_2+t_w/\triangle t}{s_1-s_2}.
\end{equation}
Thus, from Eq. (\ref{2}), 
$
\lim_{{\triangle t \rightarrow \infty} \atop{\; \mu'=t_w/\triangle t}} 
C(s_1 \triangle t+t_w,s_2 \triangle t+t_w)
$ is a function of $(s_2+\mu')/(s_1-s_2)$;
\begin{eqnarray}
& & \lim_{{\triangle t \rightarrow \infty} \atop{\; \mu'=t_w/\triangle t}} 
C(s_1\triangle t+t_w,s_2\triangle t+t_w) 
\nonumber \\
& = & \lim_{{\triangle t \rightarrow \infty} 
\atop{\; \frac{s_2+\mu'}{s_1-s_2}=\frac{s_2\triangle t+t_w}
{g[(s_1-s_2)\triangle t]}}} C(s_1\triangle t+t_w,s_2\triangle t+t_w) 
\nonumber \\ 
& = & \hat{C}(\frac{s_2+\mu'}{s_1-s_2}). 
\label{t}
\end{eqnarray}
It implies that the aging regime is the region where $t_w/\triangle t$ 
is finite. In this aging regime 
the slow-process limit of the correlation function in 
the integrand of Eq. (\ref{qs_w_a}) depends on $s_1$ and $s_2$ 
if $f(\mu)$ in Eq. (\ref{2}) is not a constant. 
It means that the work for the slow process, $W_s$, depends on the 
path of change of the external field, $h(s)$.

\item When $\lim_{\tau \rightarrow \infty}\tau/g(\tau) = 0$ and 
$\lim_{\tau \rightarrow \infty}g(\tau)/g(s \tau) \equiv g_r(s) \ 
(s < 1)$ exists, 
\begin{eqnarray}
\frac{s_2 \triangle t+t_w}{g[(s_1-s_2)\triangle t]}=
\frac{s_2 \triangle t+t_w}{g(\triangle t)}\frac{g(\triangle t)}
{g[(s_1-s_2)\triangle t]}
\nonumber \\
\rightarrow \frac{t_w}{g(\triangle t)}g_r(s_1-s_2),
\label{aging_regime}
\end{eqnarray}
as $\triangle t \rightarrow \infty$.
An example of such $g(\tau)$ is $g(\tau)={\tau}^{\alpha} \ (\alpha > 1)$. 
When $g(\tau)={\tau}^{\alpha} \ (\alpha > 1)$, 
Eq.\ (\ref{aging_regime}) holds; 
\begin{equation}
\frac{s_2\triangle t+t_w}{g[(s_1-s_2)\triangle t]}\rightarrow 
\frac{t_w}{g(\triangle t)}\frac{1}{(s_1-s_2)^{\alpha}}
\end{equation}
as $\triangle t\rightarrow\infty$.
From Eqs.\ (\ref{2}) and (\ref{aging_regime}), 
\begin{eqnarray}
& & \lim_{{\triangle t \rightarrow \infty} \atop{\; \mu'=t_w/g(\triangle t)}} 
C(s_1\triangle t+t_w,s_2\triangle t+t_w) 
\nonumber \\
 & = & 
\lim_{{\triangle t \rightarrow \infty} 
\atop{\; \mu'g_r(s_1-s_2)=\frac{s_2\triangle t+t_w}{g[(s_1-s_2)\triangle t]}
}} C(s_1\triangle t+t_w,s_2\triangle t+t_w) 
\nonumber \\
 & = & \hat{C}[\mu'g_r(s_1-s_2)]. 
\label{st}
\end{eqnarray}
It implies that 
the aging regime is the region where $t_w/g(\triangle t)$ 
is finite. In this aging regime 
the slow-process limit of the correlation function in 
the integrand of Eq. (\ref{qs_w_a}) is a function of $s_1$ and $s_2$ 
if $f(\mu)$ in Eq. (\ref{2}) is not a constant.
It means that the work $W_s$ depends on $h(s)$.

\item When $\lim_{\tau \rightarrow \infty}\tau/g(\tau) = 0$ and 
$\lim_{\tau \rightarrow \infty}g(\tau)/g(s \tau)=\infty \ 
(s < 1)$, we prove below that the aging regime for experimental time scales
does {\it not} exist. We assume that the aging regime exists. 
It means that a function $p(t)$ exists such that 
$\lim_{{\triangle t \rightarrow \infty} 
\atop{\; \mu=t_w/p(\triangle t)}}
(s_2\triangle t+t_w)/g[(s_1-s_2)\triangle t]$ is finite.
Then, from the condition $\lim_{\tau\rightarrow\infty}\tau/g(\tau)=0$, 
\begin{eqnarray}
& & \lim_{{\triangle t \rightarrow \infty} 
\atop{\; \mu=\frac{t_w}{p(\triangle t)}}}
\frac{s_2\triangle t+t_w}{g[(s_1-s_2)\triangle t]}
\nonumber \\
& = &
\lim_{{\triangle t \rightarrow \infty} 
\atop{\; \mu=\frac{t_w}{p(\triangle t)}}}
\frac{t_w}{g[(s_1-s_2)\triangle t]}
\nonumber \\
& = &
\lim_{\triangle t \rightarrow \infty}
\frac{\mu p(\triangle t)}{g[(s_1-s_2)\triangle t]}.
\end{eqnarray}
Assuming that $s'_1-s'_2 > s_1-s_2$, we see from the condition 
$\lim_{\tau\rightarrow\infty}g(\tau)/g(s \tau)=\infty \ (s<1)$ that 
\begin{eqnarray}
& & \lim_{{\triangle t \rightarrow \infty} 
\atop{\; \mu=\frac{t_w}{p(\triangle t)}}}
\frac{s_2\triangle t+t_w}{g[(s_1-s_2)\triangle t]}/
\lim_{{\triangle t \rightarrow \infty} 
\atop{\; \mu=\frac{t_w}{p(\triangle t)}}}
\frac{s'_2\triangle t+t_w}{g[(s'_1-s'_2)\triangle t]} 
\nonumber \\
 & = & 
\lim_{\triangle t \rightarrow \infty}
\frac{\mu p(\triangle t)}{g[(s_1-s_2)\triangle t]}\frac{g[(s'_1-s'_2)
\triangle t]}{\mu p(\triangle t)}
\nonumber \\  
& = &  
\lim_{\triangle t \rightarrow \infty}\frac{g[(s'_1-s'_2)\triangle t]}
{g[(s_1-s_2)\triangle t]}
\nonumber \\  
& = & \infty.
\label{mujun}
\end{eqnarray}
Since Eq. (\ref{mujun}) contradicts the assumption that 
$\lim_{{\triangle t \rightarrow \infty} 
\atop{\; \mu=t_w/p(\triangle t)}}
(s_2\triangle t+t_w)/g[(s_1-s_2)\triangle t]$ is finite, 
the aging regime for experimental time scales 
does {\it not} exist in this case. 
Hence, if only this aging regime characterized by such $g(t)$ exists, 
$\lim_{{\triangle t\rightarrow\infty}\atop{\; t_w\sim \triangle t}}
C(s_1\triangle t+t_w, s_2\triangle t+t_w)$ is equal to $0$ or $q$. 
It implies that the work is independent of $h(s)$. 
An example of such $g(t)$ is $g(t)=\exp(t)$.
\end{enumerate}

Consequently, we conclude that 
{\it if there are aging regimes characterized by $g(\tau)$ of 
case 2. or case 3. 
and the function $\hat{C}(\mu)$ is not
a constant and the FDT violation factor $X$ is not equal to $0$, 
the work for the slow process depends on the path of changing the external 
field $h(s)$. In this case, contrary to the fundamental assumption 
of thermodynamics, the work for the slow process is not the 
difference of a state function. }

\section{The work in the aging regime; results for single-correlation-scale 
systems and multi-correlation-scale systems}
\label{section5}

It has been explicitly checked on several disordered models that in the
long-time limit two situations with different dynamical behavior seem to
exist \cite{Vincent96}:
\begin{itemize}
\item There are systems with only one correlation scale apart from the
      quasiequilibrium regime, which we call ``single-correlation-scale
      systems''. For these systems, the correlation function scales
      as $C(t,t')=C[\xi(t)/\xi(t')]$ and the FDT violation factor $X$ is
      a constant.
      Equilibrium states of these systems are solved by a one
      step replica symmetry breaking ansatz. Examples are the $p$-spin
      spherical spin-glasses \cite{Cugliandolo93} and a Lennard-Jones binary
      mixture which is a model of structural
      glasses in a glassy state \cite{Barrat99}. 
      The real spin-glasses such as AgMn seem to 
      belong to this class since it
      is known that the correlation function is scaled with single scaling
      function $\xi(t)$.
\item There are systems such as Sherrington-Kirkpatrick (S-K) model which
      have an infinite number of correlation-scales apart from the
      quasiequilibrium regime \cite{Cugliandolo94}, which we call 
      ``multi-correlation-scale systems''. 
      For these systems, ultrametricity in time holds 
      for any correlation such that $C(t,t')<q$; $C(t_1,t_3)=
      \min[C(t_1,t_2),C(t_2,t_3)] \ \mbox{when} \ t_1>t_2>t_3 \ \mbox{and} \ 
      t_3\rightarrow \infty$. 
      The FDT violation factor $X$ is a nontrivial function of
      $C$. Equilibrium properties of these systems are solved by a full
      replica-symmetry breaking ansatz.
\end{itemize}

In this section, we analyze the path-dependence of the work for the slow 
process in the aging regime for
these two classes of systems by using the general results obtained in 
the previous section, Sec.\ \ref{section4}. 
It is important to note that only these 
two classes of systems seem to exist \cite{Vincent96}.

\subsection{The single-correlation-scale systems}

For the single-correlation-scale systems, scaling of the correlation
function $C(t,t')=C[\xi(t)/\xi(t')]$ holds. 
Hence, the work for the slow process in the
aging regime is given as 
\begin{eqnarray}
W_{s}/(\triangle H)^2  & = & -\int_0^1ds_1\dot{h}(s_1)
\int_0^{s_1}ds_2\dot{h}(s_2)
\nonumber \\
& & \times 
\chi\{C[\lim_{{\triangle t \rightarrow \infty} 
\atop{\; t_w \sim \triangle t}}
\frac{\xi(s_1\triangle t+t_w)}{\xi(s_2\triangle t+t_w)}]\}.
\label{w_qs_a_2}
\end{eqnarray}
Three explicit choices of the
scaling function $\xi(t)$ have been proposed so far; 
$\xi(t)=t, \xi(t)=\exp[(t/\tau_0)^{1-\alpha}/(1-\alpha)]$ and 
$\xi(t)=\exp[\ln^a(t/\tau_0)]$ \cite{Vincent96}. 
We analyze below the properties of the work for the slow process 
for each three cases.
\begin{enumerate}
\item When $\xi(t)=t$, 
the correlation function decays in power-law as found in the 
trap model \cite{Bouchaud92} and the $p$-spin 
spherical spin-glasses \cite{Cugliandolo93}. 
The slow-process limit of the correlation function in 
Eq.\ (\ref{w_qs_a_2}) depends on $s_1$ and $s_2$ as 
\begin{equation}
\lim_{{\triangle t \rightarrow \infty} 
\atop{\; \mu=\frac{t_w}{\triangle t}}} 
\frac{\xi(s_1 \triangle t+t_w)}{\xi(s_2 \triangle t+t_w)}=
\frac{s_1+\mu}{s_2+\mu}.
\end{equation}
This is an example of Eq.\ (\ref{t}) in previous section. 
\item When
      $\xi(t)=\exp[\frac{1}{1-\alpha}(\frac{t}{\tau_0})^{1-\alpha}] 
      (\alpha < 1)$ \cite{scaling}, 
the slow-process limit of 
the correlation function in Eq.\ (\ref{w_qs_a_2}) 
depends on $s_1$ and $s_2$ through the scaling function which behaves as 
\begin{equation}
\lim_{{\triangle t \rightarrow \infty} 
\atop{\; \mu = t_w/\triangle t^{1/\alpha}}}
\frac{\xi(s_1 \triangle t+t_w)}{\xi(s_2 \triangle t+t_w)}=
\exp[\frac{\mu}{\tau_0^{1-\alpha}}(s_1-s_2)].
\end{equation} 
This is an example of Eq.\ (\ref{st}).
\item When $\xi(t)=\exp [\ln^a(t/\tau_0)] (a > 1)$ \cite{scaling2},
the slow-process limit of 
the correlation function which appears in Eq.\ (\ref{w_qs_a_2}) 
depends on $s_1$ and $s_2$ 
through the scaling function which behaves as 
\begin{equation}
\lim_{{\triangle t \rightarrow \infty} 
\atop{\; \mu = \triangle t/(t_w \ln^{1-a}t_w)}}
\frac{\xi(s_1 \triangle t+t_w)}{\xi(s_2 \triangle t+t_w)}=
\exp[a \mu(s_1-s_2)].
\end{equation}
This is another example of Eq.\ (\ref{st})
\end{enumerate}
To my knowledge, only three types of the scaling function 
$\xi(t)$ discussed above are known so far 
for single-correlation-scale systems. For each of three cases 
the slow-process limit of the correlation function in 
the integrand of Eq.\ (\ref{qs_w_a}) is a function of $s_1$ and $s_2$. 
It means for the 
single-correlation-scale systems that the work for the slow process 
in the aging regime 
depends on the path of changing the external field $h(s)$ and
contrary to the assumption of thermodynamics the work 
for the slow process is not a difference of a state function.

In order to see more explicit form of the work for the slow process 
in the aging regime, we introduce the modified FDT 
Eq.\ (\ref{generalized_FDT}) with constant violation factor $X$ 
which has been verified to be valid for the single-correlation-scale 
systems \cite{Cugliandolo94}.

From Eq.\ (\ref{generalized_FDT}), the susceptibility is given by 
\begin{equation}
\chi(t,t') = -\frac{1}{k_{B}T^{eff}} C(t,t')+ \frac{q}{k_{B}T^{eff}} 
-\frac{C(0)-q}{k_{B}T},
\end{equation}
where $ C(0) \equiv \lim_{t' \rightarrow \infty} C(t',t') $. 
Hence, from Eq.\ (\ref{qs_w_a}), the work for the slow process is given by 
\begin{eqnarray}
W_{s}/(\triangle H)^2 = -\frac{C(0)-q}{2k_{B}T}- 
\frac{1}{k_{B}T^{eff}} \int_{0}^{1}ds_{1} \frac{dh(s_{1})}{ds_{1}} 
\int_{0}^{s_1}ds_{2} \frac{dh(s_{2})}{ds_{2}} \nonumber \\
\times \{q- 
\lim_{{\triangle t \rightarrow \infty} \atop{\; t_w \sim \triangle t}}
C[\xi(s_{1}\triangle t+t_{w})/\xi(s_{2}\triangle t+t_{w})]\}
\label{qs_w_a_2}.
\end{eqnarray}

Since $\lim_{{\triangle t\rightarrow\infty}\atop{\; t_w\sim\triangle t}} 
C(s_{1}\triangle t+t_{w},s_{2}\triangle t+t_{w})<q$ 
when $s_{1}\neq s_{2}$, the second term on the right hand side is non-zero. 
Hence, if $T^{eff}$ is finite the work for the slow process 
is not the difference of a state function since the work depends on the path 
of changing the field $h(s)$. In addition, the value of the work  
is different from the value of the work for the slow process in the 
quasiequilibrium regime: 
the first term on the right hand side of Eq.\ (\ref{qs_w_a_2}).
In addition, it is also shown that since the correlation function is positive 
and smaller than the dynamical E-A order parameter in this regime, 
there are bounds for the value of the quasistatic work as 
\begin{equation}
-\frac{C(0)-q}{2k_B T}-\frac{q}{2k_B T^{eff}} \leq 
\frac{W_{s}}{(\triangle H)^2} \leq -\frac{C(0)-q}{2k_B T} \label{bound}.
\end{equation}
Eq.\ (\ref{bound}) shows that the work $W_s$ in the aging regime 
is independent of the path of changing the field and 
coincides with the work $W_s$ in the quasiequilibrium regime 
when the effective temperature $T^{eff}$ is infinite, 
which holds for the $2$-spin spherical spin-glass model \cite{Cugliandolo95} 
and a model of phase separation \cite{Berthier99}.
 
\subsection{Possibility of observation of path-dependence}

In the previous subsection, 
we saw that the work for the slow process depends on 
the path of changing the external field when $\triangle t \sim t_w$. 
We show here that the path-dependence of the work for the slow process 
is strong enough to be observed in experiments by evaluating 
the path-dependence of the $5$-spin spherical spin-glass model. 

First, we describe the several pieces of information needed to evaluate 
Eq.\ (\ref{qs_w_a_2}).
It is known that the scaling of the correlation function, such 
that $C(t,t') \simeq q(t'/t)^\gamma$ and the modified FDT with the effective 
temperature, i.e. Eq.\ (\ref{generalized_FDT}), hold for the model
\cite{Cugliandolo93}. 
When $k_{B}T/J = 0.2$ \cite{gaussian}, $\gamma \simeq 0.25$ has been obtained. 
The dynamical E-A parameter $q$ and the effective temperature are given by 
$q \simeq 0.929$ and $T/T^{eff} \simeq 0.227$ when $k_{B}T/J = 0.2$. 
In computation of these values, we uses the formulae: 
\[
q^{3} (1-q)^{2} = \frac{(k_{B}T)^2}{10}, \ \ 
\frac{T}{T^{eff}}=\frac{3 (1-q)}{q}.
\]

In order to evaluate the magnitude 
of the path-dependence, we compute 
the relative difference between the works for 
two paths ($h(s)=s$ and $h(s)=s^{2}$) which is defined as 
the difference between the works 
for the two paths divided by the average of the two works. 
The results obtained from Eq.\ (\ref{qs_w_a_2}) are as follows; 
When the ratio of the measurement time to 
the waiting time is $1 : 1$, the relative difference is equal to $0.037$, 
i.e. about $4 \%$; When the ratio is $5 : 1$, the relative difference is 
equal to $0.094$, i.e. about $9 \%$. 

\begin{figure}
\centerline{\epsfxsize=10cm \epsffile{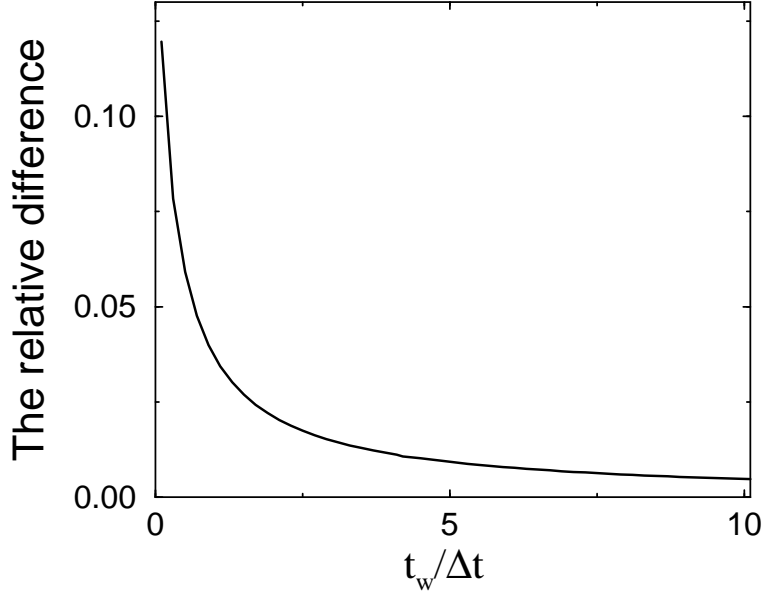}}
\caption{
The relative difference between the 
works for the two paths ($h(s)=s$ and $h(s)=s^{2}$) 
of changing the field infinitely slowly is 
plotted against the ratio of the waiting time to the measurement time 
for the $5$-spin spherical spin-glass model. 
The path-dependence of the work for the slow process in the 
aging regime is strong enough to be observed in simulations 
for the model. In addition, we see that 
the path-dependence becomes larger, as one increases the measurement time 
with keeping the waiting time fixed.
}
\label{figure1}
\end{figure}

Hence, we can say that the path-dependence of the work in the slow
process can be observed for the $5$-spin spherical spin-glass model. 
The relative difference is plotted against the ratio of the waiting time to 
the measurement time in Fig.\ \ref{figure1}. 
It tells that contrary to intuition 
the path-dependence becomes larger, as one increases the 
measurement time with keeping the waiting time fixed.

\subsection{The multi-correlation-scale systems}

For the multi-correlation-scale systems such as S-K model, it is 
known that the ultrametricity in time holds 
for all correlations such that $C(t,t')<q$ in the long-time limit 
\cite{Cugliandolo94}; 
\begin{equation}
C(t_1,t_3) = \min[C(t_1,t_2),C(t_2,t_3)] \ \ \mbox{when} \ t_1 > t_2 > t_3.
\label{ultrametric_relation_2}
\end{equation}

By using this ultrametric relation Eq.\ (\ref{ultrametric_relation_2}), 
we prove below that the correlation function in the aging regime is a
constant in the two cases 
(case 2 and case 3 in Sec.\ \ref{section4}) 
where the work for the slow process can depend on the 
path of changing the external field $h(s)$. 
Thus, for these two cases, the work for the slow process is independent of 
the path $h(s)$. 
It has been proven in Sec.\ \ref{section4} for the other two
cases that the work for the slow process cannot depend on $h(s)$. 
Thus, the work for the slow process for the multi-correlation-scale
systems is independent of the path of changing the field 
in all the experimental time regimes. 

\begin{enumerate}
\item When $g(\tau)=\tau$, by substitution 
      $t_1 \rightarrow t, t_2 \rightarrow \mu' t, t_3 \rightarrow \mu t$ in
      Eq. (\ref{ultrametric_relation_2}) and assuming that $\mu'>\mu$, we see that 
      \begin{equation}
       \lim_{t\rightarrow \infty}C(t,\mu t)=
       \min[\lim_{t\rightarrow \infty}C(t,\mu't),
       \lim_{t\rightarrow \infty}C(\mu' t,\mu t)].
       \label{ult2}
      \end{equation}
      Hence, 
      \begin{equation}
       \lim_{t\rightarrow \infty}C(\mu't,\mu t)=
       \lim_{\mu't\rightarrow\infty}C(\mu' t,\frac{\mu}{\mu'}\mu' t)=
       \check{C}(\mu/\mu')
      \end{equation}
      where $\check{C}(\mu)\equiv\lim_{t\rightarrow \infty}C(t,\mu t)$. 
      Thus, from Eq. (\ref{ult2}), 
      \begin{equation}
       \check{C}(\mu)=\min[\check{C}(\mu'),\check{C}(\mu/\mu')].
      \end{equation}
      Since it implies that 
      \begin{equation}
       \check{C}(\mu)= \left\{ \begin{array}{ll}
		      \check{C}(\mu/\mu') & \mbox{if $\mu'^2>\mu$} \\
	              \check{C}(\mu')     & \mbox{otherwise,}
		      \end{array}
       \right.
      \end{equation}
      $\check{C}(\mu)$ is a constant and hence 
      $\lim_{{\tau\rightarrow\infty} \atop{\;
      \nu=t'/\tau}}C(\tau+t',t')=\check{C}[\nu/(1+\nu)]$ is a constant.
\item When $\lim_{\tau \rightarrow \infty}\tau/g(\tau) = 0$ and 
$\lim_{\tau \rightarrow \infty}g(\tau)/g(s \tau) \equiv g_r(s) \ 
(s < 1)$ exists, the ultrametric relation Eq. (\ref{ultrametric_relation_2}) is written as 
\begin{eqnarray}
& & \lim_{{\tau\rightarrow\infty}\atop{\; \mu=t'/g(\tau)}}C(\tau+t',t')
\nonumber \\
& = &
\min[\lim_{{\tau\rightarrow\infty}\atop{\; \mu=t'/g(\tau)}}C(\tau+t',t'),
\lim_{{\tau\rightarrow\infty}\atop{\; \mu=t'/g(\tau)}}C(a\tau+t',t')]
\label{ult3}
\end{eqnarray}
where $a$ is a constant such that $0<a<1$.
The second argument of $\min$ in Eq. (\ref{ult3}) is rewritten as 
\begin{eqnarray}
\lim_{{\tau\rightarrow\infty}\atop{\; \mu=t'/g(\tau)}}C(a\tau+t',t') & = & 
\lim_{{\tau\rightarrow\infty}\atop{\; \mu=t'/g(a\tau/a)}}C(a\tau+t',t')
\nonumber \\
& = & \lim_{{\tau\rightarrow\infty}\atop{\; \mu=t'/g(\tau/a)}}C(\tau+t',t')
\nonumber \\ & = & 
\lim_{{\tau\rightarrow\infty}\atop{\; \mu=t'/[g_r(a)g(\tau)]}}C(\tau+t',t')
\nonumber \\ & = &
\lim_{{\tau\rightarrow\infty}\atop{\; \mu g_r(a)=t'/g(\tau)}}C(\tau+t',t')
\nonumber \\ & \equiv &
\bar{C}[\mu g_r(a)],
\label{cat}
\end{eqnarray}
where $\bar{C}(\mu)\equiv\lim_{{\tau\rightarrow\infty}
\atop{\mu=t'/g(\tau)}} C(\tau+t',t')$.
Similarly, the first argument of $\min$ in Eq. (\ref{ult3}) is rewritten
      as 
\begin{eqnarray}
\lim_{{\tau\rightarrow\infty}\atop{\; \mu=t'/g(\tau)}}C(\tau+t',a\tau+t')
& =  &
\lim_{{\tau\rightarrow\infty}\atop{\; \mu=(t'+a\tau)/g(\tau)}}
C(\tau+t',a\tau+t') \nonumber \\ & = & 
\lim_{{\tau\rightarrow\infty}\atop{\; \mu=t'/g(\tau)}}C[(1-a)\tau+t',t']
\nonumber \\
& = & \bar{C}[\mu g_r(1-a)]
\end{eqnarray}
where the last equality is shown from Eq. (\ref{cat}).
Consequently, 
\begin{equation}
\bar{C}(\mu)=\min\{\bar{C}[\mu g_r(1-a)],\bar{C}[\mu g_r(a)]\}.
\end{equation}
Since $g_r(a)>1$, $\bar{C}(\mu)\equiv
\lim_{{\tau\rightarrow\infty}\atop{\; \mu=t'/g(\tau)}}C(\tau+t',t')$ is
      a constant.
\end{enumerate} 

We proved above only that the correlation function is a constant within a 
correlation-scale since ultrametricity in time is valid within a 
correlation-scale.
On the other hand, even for the multi-correlation-scale systems, 
the value of the correlation function becomes $0$ when the time difference 
is much larger than the waiting time. So, value of 
the correlation function must depend on an infinite number of 
correlation-scales. 
Hence, there is possibility that there is an infinite number of experimental 
time regimes characterized by different values of the work for the slow 
process.

\section{Discussion and conclusions}
\label{section6}

Summarizing our results, for systems with aging 
we have considered the experimental time scale dependence of the 
work exerted by modulating the external field in an isothermal process.
We have shown that there are two experimental time regimes characterized by 
different behavior of the work. In the quasiequilibrium regime 
$(\triangle t \ll t_w)$, 
the work for the slow process is independent of the path of changing the
external field and the value of the work for the slow process 
is different from the change of the Helmholtz free energy obtained by 
statistical mechanics with the assumption of ergodicity. 
When the process is not infinitely slow, 
the difference between the work for non-slow process and 
the work for the slow process 
is positive for any path of changing the filed, 
i.e. the principle of minimum work holds. 
In addition, we derived the dependence of the difference 
on the measurement time from 
the long-time behaviors of the correlation function.

In the aging regime $(\triangle t \sim t_w)$, 
whose precise meaning is given in Sec. \ref{section4}, 
contrary to a fundamental assumption of thermodynamics, 
the work exerted in the slow process 
on the single-correlation-scale systems for which 
the FDT violation factor $X$ is not equal to $0$ 
depends on the path of changing the external field. 
It was examined that for 
the $5$-spin spherical spin-glass model the magnitude of the path-dependence 
is strong enough to be observed in experiments. 
The conditions when the work for the slow process depends on the path 
is also derived generally. 
On the other hand, for the multi-correlation-scale systems, 
the work for the slow process is independent of the path of changing
the field. However, the work for the slow process 
still may depend on history of a system since 
there is possibility that 
there is an infinite number of experimental time regimes characterized
by different values of the work for the slow process.
In order to attack on this problem, information on time scale of the aging
regime of multi-correlation-scale systems 
beyond time reparametrization invariance is necessary 
\cite{time-reparametrization}.

It is important to note that these results have a 
much broader range of validity, because they were derived 
without any assumption on microscopic properties of a system.

We consider the implications of our results on thermodynamics for 
systems with aging. In the quasiequilibrium regime, 
our results are the same as the consequences of thermodynamics, where 
the work for the slow process 
is independent of the path of changing the external 
field and the work for a non-slow process is 
larger than that for the slow process. Only discrepancy is difference between 
the value of the work for the slow process and 
that of the free energy obtained by 
statistical mechanics with the assumption of ergodicity. 
However, this discrepancy does not mean that 
the framework of usual thermodynamics is
invalid since the free energy can be defined by the work for the slow process. 
This suggests the usual framework of thermodynamics is valid in the 
quasiequilibrium regime even though the system is out of equilibrium. 

We considered only the properties of the work in the 
linear response regime. Thus, in order to prove validity of usual
thermodynamics in the quasiequilibrium regime, 
the relation between the heat and the 
entropy has to be considered and extension of our analysis outside the 
linear response regime has to be performed. 
For these purposes, it would be interesting to analyze Langevin dynamics of 
spin-glass models with stochastic energetics \cite{Sekimoto97,Sekimoto97-2}.

Since the work for the slow process in the quasiequilibrium regime 
is independent of history of a system in linear response regime, 
it gives a fundamental state function 
for thermodynamics in the quasiequilibrium regime and is 
worth being measured in experiments.
We discussed an experiment to measure it. 
Our results in Sec.\ \ref{section3} for the measurement time 
dependence of the work 
shows how long measurement time is needed to reach the expected accuracy. 

However, since the waiting time can be long but should be finite in reality, 
the measurement time has to be finite in order to keep the experimental time 
scale in the quasiequilibrium regime. 
Otherwise the value of the work does not come close to that of the work 
for the slow process in the quasiequilibrium regime since the
experimental time scale is in the aging regime. 
Hence, one cannot expect that the value of the measured work coincides with 
that of the work for the slow process with arbitrary accuracy. 
In other words, the accuracy of measurement of the state 
function is bounded by the length of the waiting time. 

On the other hand, in the aging regime, our results imply that 
in contrast to the usual framework of thermodynamics it is essential to 
consider the experimental time scales such as the waiting time and the 
measurement time. Furthermore, 
history of a system as a state variable is 
necessary to describe thermodynamic properties of glassy
materials. Although constructing thermodynamics for the aging regime 
is a tough work, it is a fruitful and challenging problem to be
investigated since some universal properties in the aging
regime are known \cite{Cugliandolo97}. 
Recently, a framework of thermodynamics for 
the aging regime when the temperature changes in time is proposed 
\cite{Nieuwenhuizen2000}. The framework avoids treating the problem of 
path-dependence by fixing a path to change the temperature. 
It is not clear whether thermodynamics for a single path to 
change the temperature is useful. Its validity and usefulness should be 
checked.

\acknowledgments

I wish to acknowledge valuable discussions with Takashi Odagaki, 
Akira Yoshimori, Hiizu Nakanishi, Kazuo Kitahara, Osamu Narikiyo, 
Miki Matsuo, Ken Sekimoto, Shin-ichi Sasa, Yoshihisa Miyamoto, 
Hajime Takayama, Koji Hukushima and Hitoshi Inoue.

\appendix
\section{Derivation of the long measurement time behavior of $W-W_s$}
\label{appendix}

In this appendix, we derive the long measurement time behavior of the 
deviation from the work for the slow process 
for four types of relaxation of the 
correlation function described in Sec.\ \ref{difference}.

\subsection{The case where $\lim_{t \rightarrow \infty} t [C(t)-q]$ exists}

In this section, we treat the case where 
$\lim_{t \rightarrow \infty} t [C(t)-q]$ exists, which 
includes the two cases where $\lim_{t \rightarrow \infty} t [C(t)-q]$ is 
equal to $0$ and where the limit is equal to a finite value $c$.

We rewrite the deviation given by Eq. (\ref{devi_work}) by 
introducing $K(t)$ defined as 
\begin{equation}
K(t) \equiv \int_{0}^{t} d\tau [C(\tau)-q] .
\end{equation}
The deviation is reduced to 
\begin{eqnarray}
\frac{k_B T}{(\triangle H)^2}(W-W_{qs}) = \frac{1}{\triangle t} 
\int_{0}^{1}ds_1 \frac{dh(s_1)}{ds_1} [h'(0)K(s_1 \triangle t) 
\nonumber \\
+ \int_{1}^{s_1}ds_2 h''(s_1-s_2) K(s_2 \triangle t)] \label{app:dev1}.
\end{eqnarray}

At first, we consider the deviation when 
$\lim_{t \rightarrow \infty} t [C(t)-q]$ is equal to $0$. 
Here, in order to prove that the deviation is proportional to the 
inverse of the measurement time $\triangle t$, we show that the 
proportional constant is finite. 
The proportional constant is given as 
\begin{eqnarray}
\lim_{\triangle t \rightarrow \infty}\triangle t \frac{k_B T}{(\triangle H)^2}
(W-W_{qs}) = \lim_{\triangle t \rightarrow \infty} 
\int_0^1 ds_1 \frac{dh(s_1)}{ds_1} [h'(0)K(s_1 \triangle t) \nonumber \\
+\int_0^{s_1}ds_2 h''(s_1-s_2)K(s_2 \triangle t)].
\end{eqnarray}
Since $\lim_{t \rightarrow \infty}K(t)$ exists in the case, $K(t)$ 
is finite. Hence, the order of the limit $\triangle t \rightarrow \infty$
and the integration can be changed. 
Thus, the proportional constant is given by 
\begin{equation}
\lim_{\triangle t \rightarrow \infty}\triangle t \frac{k_B T}{(\triangle H)^2} 
(W-W_{qs}) = K_{\infty} \int_0^1 ds \left|\frac{dh(s)}{ds}\right|^2, 
\end{equation}
where $K_{\infty}$ is defined by Eq. (\ref{k}).
Consequently, it is proved that the proportional constant is finite and 
the deviation is proportional to $1/\triangle t$.

Next, we consider the deviation when 
$\lim_{t \rightarrow \infty} t [C(t)-q]$ is equal to a finite value $c$.
In order to prove that the deviation is proportional to 
$\log \triangle t / \triangle t$, we show that the proportional constant 
is finite. 
Dividing the region of integration of Eq. (\ref{app:dev1}) into 
five regions, the proportional constant is given by 
\begin{eqnarray}
& & 
\lim_{\triangle t \rightarrow \infty} \frac{\triangle t}{\log \triangle t} 
\frac{k_B T(W-W_{qs})}{(\triangle H)^2}\nonumber \\
& = & 
\lim_{\triangle t \rightarrow \infty} \frac{1}{\log \triangle t} 
[h'(0)\int_0^{\frac{1}{\triangle t}}ds_1 h'(s_1) K(s_1 \triangle t) 
\nonumber \\ & + & 
h'(0)\int^1_{\frac{1}{\triangle t}}ds_1 h'(s_1) K(s_1 \triangle t) 
\nonumber \\ & + & 
\int_0^{\frac{1}{\triangle t}}ds_1 h'(s_1) \int_0^{s_1} ds_2 
h''(s_1-s_2) K(s_2 \triangle t) 
\nonumber \\ & + & 
\int^1_{\frac{1}{\triangle t}}ds_1 h'(s_1) \int_0^{\frac{1}{\triangle t}} ds_2 
h''(s_1-s_2) K(s_2 \triangle t) 
\nonumber \\ & + & 
\int^1_{\frac{1}{\triangle t}}ds_1 h'(s_1) \int_{\frac{1}{\triangle t}}^{s_1} 
ds_2 h''(s_1-s_2) K(s_2 \triangle t)] \label{app:dev2}.
\end{eqnarray}

We discuss the behavior of the function $K(s \triangle t)$ in each region. 
When $s \leq 1/ \triangle t$, from the definition of $K(t)$, Eq. (\ref{k}), 
it is shown that 
\begin{equation}
K(s \triangle t) \leq \int_0^1 d\tau [C(\tau)-q].
\end{equation}
Thus, $K(s \triangle t)$ is finite even when $\triangle t \rightarrow \infty$.
On the other hand, when $s > 1/ \triangle t$, 
we express the function $K(s \triangle t)$ by using a function defined as 
$C(t)-q \equiv c/t+f(t), t \geq 1$: 
\begin{eqnarray}
K(s \triangle t) & = & \int_0^{s \triangle t}d\tau [C(\tau)-q] 
\nonumber \\ 
& = & \int_0^1 d\tau [C(\tau)-q]+\int_1^{s \triangle t} d\tau 
[\frac{c}{\tau}+f(\tau)] 
\nonumber \\ 
& = & \int_0^1 d\tau [C(\tau)-q]+c \log s+c \log \triangle t+
\int_1^{s \triangle t} d\tau f(\tau). 
\end{eqnarray}
Since $\lim_{t \rightarrow \infty}t f(t) = 0$, the fourth term on the right 
hand side is finite even when $\triangle t$ is infinite. 
Thus, the behavior of the function is given as $K(s \triangle t) \simeq 
c \log \triangle t$ when $s > 1/ \triangle t$. 
By evaluating the each term of Eq. (\ref{app:dev2}) with using the 
above behavior of the function $K(s \triangle t)$, 
it is straightforward to show that 
\begin{equation}
\lim_{\triangle t \rightarrow \infty} \frac{\triangle t}{\log \triangle t} 
\frac{k_B T}{(\triangle H)^2} (W-W_{qs}) = c \int_0^1 ds 
|\frac{dh(s)}{ds}|^2.
\end{equation}
Consequently, it is proved that the proportional constant is finite.

\subsection{The case where $\lim_{t \rightarrow \infty} t [C(t)-q]$ does not 
exist}

In this section, we treat the case where 
$\lim_{t \rightarrow \infty} t [C(t)-q]$ does not exist, which 
includes the two types of long-time behavior of the correlation function, 
such that 
$C(\tau) \simeq q+c/\tau^{\alpha}, 0 < \alpha < 1$ and 
$C(\tau) \simeq q+c/\log \tau$. 

In order to find out the leading term in the behavior for 
the long measurement time, 
we divide the region of integration of Eq. (\ref{devi_work}) into 
three regions as 
\begin{eqnarray}
\int_0^1 ds_1 \int_0^{s_1} ds_2 = 
\int_0^{(\triangle t)^{-a}}ds_1 \int_0^{s_1}ds_2 + 
\int_{(\triangle t)^{-a}}^1 ds_1 \int_{s_1-(\triangle t)^{-a}}^{s_1} ds_2  
\nonumber \\
+\int_{(\triangle t)^{-a}}^{1} ds_1 \int_0^{s_1-(\triangle t)^{-a}} ds_2 
\label{division}, 
\end{eqnarray} 
where we assume that the parameter $a$ satisfies $0<a<1$.
Since all the integrands (the time derivative of the field and the 
correlation function) are finite, it can be shown that the 
first two terms on the right hand side are bounded as 
\[
|\mbox{the first term}| \leq \mbox{const} \times (\triangle t)^{-2 a} 
\]
and 
\[
|\mbox{the second term}| \leq \mbox{const} \times (\triangle t)^{-a} 
[1-(\triangle t)^{-a}] .
\]
In addition, we notice that since $s_2 \leq s_1-(\triangle t)^{-a}$ 
in the region of integration of 
the third term, $(s_1-s_2)\triangle t$ is large when the measurement time 
$\triangle t$ is large. Thus, the third term is determined by the 
long-time behavior of the correlation function.

At first, we treat the case where the correlation function behaves at a 
long time as $C(\tau) \simeq q+c/\tau^{-\alpha}, 0<\alpha<1$.
We divide the region of integration of the third term of Eq. (\ref{division}) 
into three regions as 
\begin{eqnarray}
\int_{(\triangle t)^{-a}}^1 ds_1 \int_0^{s_1-(\triangle t)^{-a}} ds_2 = 
\int_0^1 ds_1 \int_0^{s_1} ds_2 - 
\int_0^{(\triangle t)^{-a}} ds_1 \int_0^{s_1} ds_2 \nonumber \\ 
-\int_{(\triangle)^{-a}}^1 ds_1 \int_{s_1-(\triangle t)^{-a}}^{s_1} ds_2.
\label{division2}
\end{eqnarray}
Since all the integrands are finite, it is shown that the 
last two terms are bounded by 
\[
|\mbox{the second term}| \leq \mbox{const} \times 
(\triangle t)^{-a (2-\alpha)-\alpha}
\]
and 
\[
|\mbox{the third term}| \leq \mbox{const} \times 
(\triangle t)^{-\alpha-a(1-\alpha)}.
\]

On the other hand, by using the long-time behavior of the correlation 
function, the first term is given as 
\begin{equation}
\int_0^1 ds_1 \frac{dh(s_1)}{ds_1} \int_0^{s_1} ds_2 \frac{dh(s_2)}{ds_2} 
\frac{c}{(s_1-s_2)^{\alpha}} \frac{1}{(\triangle t)^{\alpha}} 
\label{power1}.
\end{equation}
Thus, we find that this term is $O[(\triangle t)^{-\alpha}]$. 
Assuming the free parameter $a$ is larger than the exponent $\alpha$, 
one sees that the leading term is Eq. (\ref{power1}) by comparing 
the orders of all the terms in Eqs. (\ref{division}) and (\ref{division2}). 
Hence, the long measurement time behavior of 
the deviation is given by Eq. (\ref{power}). 

Next, we consider the case where the long-time behavior of the correlation 
function is given by $C(\tau) \simeq q + c/\log \tau$. 
Again, we start by evaluating the third term on the right hand side 
of Eq. (\ref{division}).
By using the long-time behavior of the correlation function, 
the term is given by 
\begin{equation}
\int_{(\triangle t)^{-a}}^1 ds_1 \frac{dh(s_1)}{ds_1} 
\int_0^{s_1-(\triangle t)^{-a}} ds_2 \frac{dh(s_2)}{ds_2} 
\frac{c}{\log (s_1-s_2)+\log \triangle t}.
\end{equation}
Since in the region of integration the following inequality holds: 
\[
\left| \frac{\log (s_1-s_2)}{\log \triangle t} \right| \leq a,
\]
by neglecting numbers of $O(a/\log \triangle t)$ the third term of 
the right hand side of Eq. (\ref{division}) is reduced to 
\begin{equation}
\int_{(\triangle t)^{-a}}^1 ds_1 \frac{dh(s_1)}{ds_1} 
\int_0^{s_1-(\triangle t)^{-a}} ds_2 \frac{dh(s_2)}{ds_2} 
\frac{c}{\log \triangle t}.
\end{equation}
Furthermore, by neglecting numbers of 
$O[(\triangle t)^{-a}/\log \triangle t]$, the term is reduced to 
\begin{equation}
\int_0^1 ds_1 \frac{dh(s_1)}{ds_1} \int_0^{s_1} ds_2 \frac{dh(s_2)}{ds_2} 
\frac{c}{\log \triangle t} = \frac{c}{2 \log \triangle t}.
\end{equation}

Consequently, one sees that the leading term is the above one by comparing 
the orders of all the terms in Eqs. (\ref{division}) and (\ref{division2}). 
Thus, the long measurement time behavior of 
the deviation is given by Eq. (\ref{log}). 

Finally, in order to determine the order of the neglected term 
we determine the value of the free parameter $a$, 
since the neglected term contain the free parameter $a$. 
Since in terms of $a$ the order of the neglected terms are given as 
$O(a/\log \triangle t)$ and $O[(\triangle t)^{-a}]$, 
we can determine the value of $a$ so that the value of the neglected terms are 
minimized. Thus, assuming that the neglected term is given by 
$\alpha a/\log \triangle t + \beta (\triangle t)^{-a}$ 
where $\alpha$ and $\beta$ are constants, the value of $a$ is determined by 
\begin{equation}
\frac{d}{da} [\alpha \frac{a}{\log \triangle t}+\beta (\triangle t)^{-a}] 
= 0.
\end{equation}
Since the solution is $a \simeq 2 \log(\log \triangle t)/\log \triangle t$, 
one sees that the order of neglected term is 
$O[\log (\log \triangle t)/(\log \triangle t)^2]$ as shown in 
Eq. (\ref{log}).

\narrowtext


\begin{references}
\bibitem{FDT} The reason why the usual FDT holds in the quasiequilibrium 
regime though a system is out-of-equilibrium is that the change of the 
free energy stops at the long time limit, since the FDT violation is 
bounded by the time-derivative of the free energy \cite{Cugliandolo97-2}.
\bibitem{time-difference} When $s_{1}=s_{2}$, 
$(s_{1}-s_{2})\triangle t \sim s_{2}\triangle t+t_{w}$ 
does not hold even if $\triangle t \sim t_{w}$ since 
the time difference $(s_1-s_2)\triangle t$ is zero. However, the case can 
be ignored, since the susceptibility is finite there and the point 
$s_{1}=s_{2}$ does not contribute to the value of the integral by itself.
\bibitem{complementality} It has been shown from the massless Langevin 
equation without the assumption of a linear response that $W-W_{qs} \propto 
1/\triangle t$ \cite{Sekimoto97}. Since the correlation function derived 
from the massless Langevin equation obeys the exponential relaxation, 
our result agrees with that of Ref.\ \cite{Sekimoto97}.
\bibitem{scaling} This form of the scaling function was proposed to account 
for the results of experiments in polymer glasses \cite{Struik78}, 
then used in analyses of aging effects in decay of the thermoremanent 
magnetization \cite{Ocio85}, and found in the exact solution of the asymmetric 
spherical SK model \cite{Franz94}.
\bibitem{scaling2} This form of the scaling function is suggested by 
the numerical data from the analysis of a point particle in a random 
potential with infinite dimension \cite{Cugliandolo96-2} and is used to 
scale data for the thermoremanent magnetization and the out-of-phase 
susceptibility.
\bibitem{gaussian} The energy is scaled with a constant $J$ which determines 
the variance of the coupling constant between spins as 
\[
\rho(J_{i_1\cdots i_5})=\sqrt{\frac{N^4}{5! \pi J^2}} 
\exp\left( -\frac{N^4 {J_{i_1\cdots i_5}}^2}{5! J^2} \right).
\]
\bibitem{time-reparametrization} 
The equations of motion of $C(t,t')$ and $R(t,t')$ of the 
multi-correlation-scale systems in the long waiting time limit have an 
infinite set of invariance called ``time reparametrization invariance'' 
\cite{Cugliandolo94}; 
If we perform an arbitrary reparametrization of time 
\[
\hat{t}=\xi(t) \ \ \hat{t}'=\xi(t')
\]
and we redefine the two-time quantities as 
\[
\hat{C}(\hat{t},\hat{t}')=C[\xi(t),\xi(t')] \ \ 
\hat{G}(\hat{t},\hat{t}')=G[\xi(t),\xi(t')].
\]
Then, the redefined functions $\hat{C}$ and $\hat{G}$ satisfy the 
same equations of motion. It is important to note that 
this invariance is a consequence of having neglected the 
time derivatives in taking the long waiting time limit. 
The full equations of motion have no such invariance and the 
solution is unique.
\end{references}
\end{document}